\PassOptionsToPackage{numbers,sort&compress}{natbib}
\documentclass[onecolumn, groupedaddress, showkeys]{revtex4}
\usepackage{setspace}
\usepackage[dvips]{color}
\usepackage{xcolor}
\usepackage{amsfonts}
\usepackage{amsmath}
\usepackage{amssymb}
\usepackage{graphicx}
\usepackage{mathrsfs}
\usepackage{bbm}
\usepackage{bm}
\usepackage{multirow}

\begin{document}

\newcommand{\Secref}[1]{Sec.~\ref{#1}}
\newcommand{\dd}{d}
\newcommand{\pd}{\partial}
\newcommand{\myU}{\mathcal{U}}
\newcommand{\myr}{q}
\newcommand{\Urho}{U_{\rho}}
\newcommand{\myalpha}{\alpha_*}
\newcommand{\bd}[1]{\mathbf{#1}}
\newcommand{\Eq}[1]{Eq.~(\ref{#1})}
\newcommand{\Eqn}[1]{Eq.~(\ref{#1})}
\newcommand{\Eqns}[1]{Eqns.~(\ref{#1})}
\newcommand{\Figref}[1]{Fig.~\ref{#1}}
\newtheorem{theorem}{Theorem}
\newcommand{\me}{\textrm{m}_{\textrm{e}}}
\newcommand{\sgn}{\textrm{sign}}
\newcommand*{\bfrac}[2]{\genfrac{\lbrace}{\rbrace}{0pt}{}{#1}{#2}}

\renewcommand{\thesection}{\arabic{section}}
\renewcommand{\thesubsection}{\thesection.\arabic{subsection}}
\renewcommand{\thesubsubsection}{\thesubsection.\arabic{subsubsection}}


\title{Cosmological mass of the photon related to Stueckelberg and Higgs mechanisms}

\author{Lorenzo Gallerani Resca} 
\email{resca@cua.edu}
\homepage{http://physics.cua.edu/people/faculty/homepage.cfm} 

\affiliation{Department of Physics and Vitreous State Laboratory, 
The Catholic University of America,  
Washington, DC 20064}

\date{\today} 

\begin{abstract}

\textbf{ABSTRACT} --- I consider electro-weak (EW) masses and interactions generated for photons by vacuum expectation values of Stueckelberg and Higgs fields. I provide a prescription to relate their parametric values to a cosmological range derived from a fundamental Heisenberg uncertainty principle and Einstein-de Sitter cosmological constant and horizon. This yields qualitative connections between microscopic ranges acquired by $W^{\pm}$ or $Z^0$ gauge Bosons and the cosmological scale and minimal mass acquired by $g$-photons. I apply that procedure to an established Stueckelberg-Higgs mechanism, while I consider a similar procedure for a pair of Higgs fields that may spontaneously break all U(1)xSU(2) gauge invariances. My estimates of photon masses and their additional parity-breaking interactions with leptons and neutrinos may be detectable in suitable accelerator experiments. Their effects may also be observable astronomically through massive $g$-photon condensates that may contribute to dark matter and dark energy.

\end{abstract}

\keywords{Stueckelberg mechanisms; Higgs mechanisms; photon mass; neutrino mass; cosmological constant; vacuum energy; dark energy; dark matter; Bose-Einstein condensate; general theory of relativity. --- \textbf{EMAIL}: resca@cua.edu} 

\maketitle

\section{Introduction}\label{Introduction} 


\maketitle

In recent papers I proposed that photons, as quantum mechanical (QM) particles, must carry a minimal cosmological mass due to causal self-connection and entanglement within a cosmological horizon.\cite{RescaPhoton,resca2023cosmological} Such photon mass, $m_g$, must  be of the order of    
\begin{equation}\label{massfundamental}
M_g c^2 = \frac{h c}{L_g} = h c \sqrt{\Lambda} \simeq 1.3 \mathrm{x}10^{-41} GeV ,
\end{equation}  
where its corresponding Compton wavelength, 
\begin{equation}\label{fundamental} 
L_g = \sqrt{\frac{1}{\Lambda}} \simeq 10^{10} \mathrm{ly} \simeq 0.946 \mathrm{x} 10^{26} m, 
\end{equation}  
is the inverse square-root of Einstein's cosmological constant, $\Lambda$, which is related in turn to de Sitter horizon. Namely, \Eq{massfundamental} is fundamentally based on the Uncertainty Principle in the relativistic limit, while \Eq{fundamental} provides the most basic cosmological delocalization range and expansion of the de Sitter Space. Together with a cosmological time, $T_g = L_g / c \simeq 10^{10} \mathrm{years} \simeq 3.2 \mathrm{x} 10^{17} s$, I have defined and used a corresponding set of $g$-units with further objectives.\cite{RescaPhoton, resca2023cosmological} 

Other authors have reached similar, if not equivalent conclusions. Remarkably, using a far more sophisticated logotropic model of universe expansion,\cite{ChavanisLogotropic, ChavanisCoreHalo} Chavanis has arrived to mass equations virtually identical to mine. The main, if not the only difference, is that Chavanis does not associate his fundamental dark-fluid particle, or `cosmon,' to any more specific elementary particle within or beyond the Standard Model (SM),\cite{Nir} as I did with regard to $g$-photons. Capozziello and colleagues have referred to photons more specifically,\cite{Capozziello} but they maintained expressions such as \Eq{massfundamental} as lowest bounds of uncertainty, rather than estimates of an actual and finite photon mass, as I assumed. Hui and colleagues have deeply and extensively referred to Bose-Einstein condensates (BEC) that may provide major components to dark matter (DM) and dark energy (DE),\cite{Hui} as I conceived independently, attributing DE and components of DM to $g$-photons.\cite{RescaPhoton}

Presently it is generally believed that QM supersedes classical mechanics (CM) at all scales. Thus, `exact' QM equations for electro-magnetic (EM) fields should directly involve EM potentials, $A^{\mu}$, as well as Planck constant, $h$. That extends the approximation of the purely classical Maxwell equations to include at least some Proca-like mass terms and a corresponding Compton wavelength for the photon. It is further envisioned that QM may eventually merge with general relativity (GR) to form a more encompassing unified field theory. \Eq{massfundamental} may provisionally work as a low energy limit in that perspective. 

Now, all known elementary particles have Compton extensions confined to a microscopic range, while the photon extends over either an infinite range, classically, or a cosmological scale, as I and others have considered. In this paper I will inquire on how that range and possible mass of the photon may derive from QM interactions with other elementary particles of the SM. The basic QM mechanism that I will consider is spontaneous symmetry breaking (SSB) of vacuum expectation values (VEV) of certain quantized fields. The primary theory to which I refer has to be that of electro-weak (EW) interactions developed by Weinberg and Salam (WS).\cite{Mandl, peskin1995introduction} That is a gauge theory with U(1)xSU(2) symmetries, spontaneously broken with a weak isospin doublet Higgs field, $\mathbf \Phi(x)$. However, EW-WS theory requires from the start that the photon be massless and thus that electric charge be exactly conserved. Mathematically, that requires that an Abelian subgroup, $U(1)_{EM}$, of $U(1)_Y \mathrm{x} SU(2)_L \rightarrow U(1)_{EM}$, remains invariant under gauge transformations, exactly. That is where I will differ, ever so slightly, but with fundamental consequences nonetheless.

There are basically two approaches that may allow the photon to attain some type of mass through further, hence complete, SSB. The first one depends on combining a Stueckelberg mechanism with the standard EW-WS Higgs mechanism. A vast literature has developed around that field, way beyond Stueckelberg's original considerations.\cite{Ruegg} A second, either equivalent or even more promising approach, is to combine the standard EW-WS Higgs mechanism with another kind of Higgs field. In this paper I will suggest templates for both approaches and their relations, including my fundamental `prescription' that limits the mass of the photon to its minimal cosmological bound, as reported in \Eq{massfundamental}: for a deeper discussion about that, see Ref. \onlinecite{resca2023cosmological} in particular.

\section{A cosmological Stueckelberg - Higgs mechanism}\label{Stueckelberg}

A minimal requirement to develop any sensible EW theory is to consider some SSB of an initially gauge invariant U(1)xSU(2) symmetry group, as discussed  on pp. 717-719 of Ref. \onlinecite{peskin1995introduction}, for example. A most readily accessible paper and theory of a Stueckelberg-Higgs mechanism based on that requirement has been developed in a paper (KMcK) by Kuzmin and McKeon.\cite{Kuzmin:2001pg}

In KMcK theory a Stueckelberg mass term is generated for a $U(1)$ gauge field, without disrupting the SU(2) Higgs sector of the EW-WS theory. Thus the photon field, $A^\mu$, acquires a mass, $m_A$, directly related to a Stueckelberg mass, $m_S$, while still maintaining both unitarity and renormalizability of the Abelian theory. Consequently, this massive photon field, $A^\mu$, may now interact directly with neutrino fields, and it breaks parity in its interactions with electron fields. These are truly remarkable results, which can in principle provide fundamentally testable predictions. 

Based on the Stueckelberg mass, $m_S$, the basic parameter of KMcK theory is 
\begin{equation}\label{epsilon}
\epsilon = \frac{2 m_S c^2}{v} = \frac{m_S}{m_W} g . 
\end{equation}  
In \Eq{epsilon}, $v$ is the VEV of the Higgs field, $m_W  \simeq 80.4 GeV/c^2$ is the mass of the $W^{\pm}$ gauge Bosons, and $g = e / sin \theta_W$ is related to the charge of the electron, $e$, as in EW-WS theory, where $sin^2 \theta_W \simeq 0.231$. Though non-zero, exactly, $\epsilon/g$ must be very small, experimentally. This may have prompted KMcK to conclude that the Stueckelberg mass, $m_S$, should be `tuned to zero,' and that it is likely that this is a consequence of SU(2)×U(1) being a subgroup of a larger, more fundamental grand unified non-Abelian gauge group, which is spontaneously broken to SU(2)×U(1).\cite{Kuzmin:2001pg} Similar considerations have been developed independently in string theories.\cite{Reece, Axiverse}

Actually, continuously `tuning to zero' fundamental constants that have physical dimensions is a `risky' proposition. For example, Planck constant, $h = 6.62607015 \mathrm{x} 10^{-34} J/{Hz}$, is exceedingly small on macroscopic scales and units. Yet, `tuning' $h \rightarrow 0$ causes all of quantum statistics and QM to vanquish. Planck length, $l_P$, mass, $m_P$, and time, $t_P$, consequently vanquish. Ditto for a `real' photon that has an invariant $g$-mass of the order of $M_g$ in \Eq{massfundamental}, because that \Eq{massfundamental} is precisely the consequence of $h$ and $c$ and $\Lambda$ being all finite on their dimensional grounds. None of those fundamental constants can be taken exactly to either zero or infinity, physically.

Now, in any quantum field theory (QFT) based on flat Minkowski space-time some basic particle masses may only enter as free parameters. In KMcK theory that is also the case for their Stueckelberg mass, $m_S$. Anything may change, however, for a QFT based on a curved space-time. Unfortunately, there is no definitive such theory as yet.\cite{BirrellCurved, WaldCurved, WaldHolland, Ford} In fact, rigorous definitions of particles may not even be possible on curved space-times. It is believed that the mass of a particle becomes ambiguous if its size becomes of the order of the background curvature scale. So, my \Eq{massfundamental} has to be regarded as a preliminary estimate at best. Nevertheless, I have already discussed that \Eq{massfundamental} has a solid foundation on fundamental Heisenberg QM uncertainty and Einstein-de Sitter cosmological constant principles.\cite{RescaPhoton,resca2023cosmological}

Let me then proceed to make the main `prescription' of this paper: `When confronted with the question of replacing an exactly zero mass for the photon with a QFT non-zero free parameter, provisionally introduce the elementary estimate of \Eq{massfundamental}.'

For KMcK theory, I thus `prescribe' that the mass, $m_A$, of the photon field, $A^{\mu}$, is of the order of 
\begin{equation}\label{mass}
M_g \sim m_A = m_S cos \theta_W ,
\end{equation}  
where the second equality is derived from Eqs. (16) of KMcK. I can thus estimate that
\begin{equation}\label{epsilonmass}
\epsilon/g = \frac{m_S}{m_W} \sim \frac{M_g}{cos\theta_W m_W}\simeq 2.1 \mathrm{x}10^{-43} . 
\end{equation}  
Though exceedingly small, $\epsilon/g$ is still finite and precisely defined, at least within an order of magnitude or so. 

Now one may go on to derive quantitative estimates for all exceptional physical quantities and effects predicted by KMcK theory. Clearly, all such corrections to standard EW-WS theory are bound to be extremely small. Their detection in collider experiments may thus be thought as exceedingly difficult, if not impossible.

Let us recall, however, that a finite mass, $m_g > 0$, of $g$-photons, no matter how small, confers to them the possibility of standing at rest in a BEC. In fact, on a cosmological scale, the number density of a BEC of $g$-photons is huge,\cite{RescaPhoton,resca2023cosmological} i.e., 
\begin{equation}\label{numberdensity}
n_g = \frac{\rho_{\Lambda}} {m_g c^2} \sim \frac{c^3 \sqrt{\Lambda}} {8 \pi h G} = \frac{\sqrt{\Lambda}} {8 \pi l_P^2} \simeq 2.6 \mathrm{x}10^{41} m^{-3} ,
\end{equation}  
which is about $6 \mathrm{x} 10^{32}$ times greater than the current CMB photon density. Holding in a collider a vast reservoir of such high-density BEC of $g$-photons may eventually allow some signature detection of exceptional KMcK effects, over long times and an immense number of scattering events with other photons and particles.

Consider, for example, basic neutrino-electron scattering as depicted in two graphs of Fig. [19.6 (a,b)], p. 431, of Ref. \onlinecite{Mandl}, exchanging one $Z^0$ or one $H$ neutral Bosons. Now, in KMcK theory, a third basic diagram must be added to those, representing the exchange of a single massive $g$-photon line between those two lepton propagators. KMcK derive and state that unequivocally as a result of their conclusive Eq. [25]. Though exceedingly small, the $g$-photon vertex contribution with neutrinos may eventually be detected, thus providing an experimental confirmation of KMcK theory supplemented with my cosmological `prescription.'

Historically, experiments involving neutrino-electron scattering with the exchange of a $Z^0$ Boson were critical to confirm EW-WS theory in 1983: see discussion on p. 434, culminating with  Eq. [19.50], in Ref. \onlinecite{Mandl}. Since then, flux and detection of neutrino-electron scattering have improved drastically.\cite{Tomalak, Marshall} Perhaps some events involving a direct interaction of a neutrino with a single $g$-photon may be conclusively detected or inferred. In fact, historically, who would have thought that neutrinos, and then one after another particle predicted by the SM, could eventually be detected experimentally? But for $g$-photons that may still be asking too much. Unfortunately, not everything that can be measured is important and not everything that is important can be measured.  

From an astronomical perspective, however, the low-energy cross section of photon-photon scattering yields an estimate of mean-free-path for visible light in the CMB frame of about $7 \mathrm{x} 10^{52} \mathrm{ly}$, at least $10^{42}$ times greater than the size of the `observable' universe.\cite{Liang} Thus my $g$-BEC should still remain largely transparent to visible light, by at least a $10^{9}$ scale factor. However, clumping as dark matter in and out of the $g$-BEC may cause some dimming or gravitational lensing of most distant objects that can also be theoretically predicted and astronomically observed.\cite{resca2023cosmological}


\section{A cosmological Higgs - Higgs coupling}\label{Higgs}

An alternative procedure to arrive to an actual mass for the photon is to introduce directly in EW-WS theory an interaction with another complex scalar Higgs field, $\Psi_{g-EW}(x)$, thus causing SSB of any remaining U(1) gauge invariance. That is in addition to the weak isospin doublet Higgs field, $\mathbf \Phi(x)$, which causes SSB of the SU(2) gauge invariance at first. Together, both fields still satisfy all requirements of the Higgs Sector of U(1)xSU(2), as examined on pp. 717-719 of Ref. \onlinecite{peskin1995introduction}.

The electro-weak mass for the photon, $m_g$, should be related to VEV's of both Higgs fields, but mostly, if not exclusively, to the additional singlet Higgs field, $\Psi_{g-EW}(x)$. In fact, that should yield a parameter closely related to $\epsilon$ of \Eq{epsilon}, referring to KMcK theory. Although complicated, relations between Higgs and Stueckelberg fields are in fact well understood.\cite{Kuzmin:2001pg, Ruegg} 

In addition to all parameters of the standard EW-WS theory, the fully massive $g$-EW theory shall contain at least one additional parameter, i.e., the mass, $m_g > 0$, acquired by the fourth gauge Boson, which should be identified with the photon. To that $m_g$ parameter one may then apply my cosmological `prescription' or estimate, as provided in \Eq{massfundamental} or exemplified in \Eq{mass}. One may thus proceed to relate parametric values of VEV's to precise, although qualitative, connections between microscopic ranges acquired by $W^{\pm}$ or $Z^0$ gauge Bosons and a cosmological range, such as $L_g$, for the correspondingly acquired mass, $m_g \sim M_g$, of the fully SSB $g$-photon.

One may also recall that lepton masses, $m_l$, and those of their corresponding neutrinos, $m_{\nu_l}$, may also enter as parameters in EW theories. Basically, those parameters may be proportional to products of VEV's and Yukawa coupling constants: see Eqs. [19.8], p. 422, of Ref. \onlinecite{Mandl}, for example.

Heuristically and independently, I found that neutrino masses appear to be related to the same cosmological construct and $g$-units as
\begin{equation}\label{geometric}
\sqrt {M_g m_P} = \bigg [ \frac{h^3 \Lambda}{c G} \bigg ]^{1/4} \simeq 2 \mathrm{x} 10^{-2} e V / c^2 \simeq m_{\nu} ,
\end{equation}  
where $G$ represents Newton's gravitational constant and $m_P$ is Planck mass.\cite{resca2023cosmological} That neutrino mass was more theoretically derived in a logotropic model of the universe evolution.\cite{ChavanisCoreHalo} A similar construct for the electron mass in that logotropic model yields
\begin{equation}\label{massfinestructureChavanis}
m_e = 1.03 \alpha \bigg (\frac{\Lambda \hbar^4}{G^2} \bigg )^{(1/6)} \simeq 0.511MeV/c^2 , 
\end{equation}  
where $\alpha \simeq 1 / 137.04$ is the fine structure constant.\cite{ChavanisLogotropic}

One may then apply to a fully SSB EW-WS-Yukawa theory the same cosmological `prescription' to estimate lepton and neutrino masses, thus gaining deeper insight into \Eq{geometric} or \Eq{massfinestructureChavanis} and their possible cosmological connection to $m_g$. 

Providing a rigorous fully massive $g$-EW theory is clearly beyond the scope of this paper. However, I may at least conceive of a lowest order approximation to it, namely that of considering the ordinary isospin doublet Higgs field, $\mathbf \Phi(x)$, as completely decoupled from the additional singlet Higgs field, $\Psi_{g-EW}(x)$. Thus only the latter acts to provide a far smaller mass to the fourth $g$-Boson, the photon, through SSB of the previously remaining $U(1)_{EM}$ Abelian gauge invariance.

Let me thus provide some basic results of the standard EW-WS theory as derived in Secs. 18.3 and 19.1 of Ref. \onlinecite{Mandl}, after having restored most general units, including explicitly $c$ and $\hbar$ fundamental constants. Using those established notations and experimental results I find that
\begin{equation}\label{EWVacuum}
v =  \sqrt \frac{- 2 \mu^2 c^4} {2 \hbar c \lambda} \simeq 1.8 \mathrm{x} 10^{9} \sqrt{GeV/cm}   , 
\end{equation}  
\begin{equation}\label{EWWS}
\sqrt{-2 \mu^2} = m_H \simeq 125 GeV/c^2 ,
\end{equation}  
where $v$ is the vacuum expectation value (VEV) of the WS isospin doublet Higgs field, $\mathbf \Phi(x)$, and $m_H$ is the current experimental estimate of the corresponding EW Higgs Boson mass.

Then
\begin{equation}\label{EWLambda}
\lambda = \frac{- 2 \mu^2 c^4}{2 \hbar c v^2} \simeq 0.12 , 
\end{equation}  
\begin{equation}\label{EWW}
m_W c^2= \frac{|e| v} {2 sin \theta_W}  \simeq 80 GeV   , 
\end{equation}  
where $|e| = (4 \pi \alpha \hbar c)^{1/2} \simeq 4.3 \mathrm{x}10^{-8 } (GeV \mathrm{x} cm)^{1/2} $ is measured in `rationalized' units of Mandl \textit{et al.} and $\hbar c \simeq 1.973 \mathrm{x}10^{-14} GeV \mathrm{x} cm$. That yields the $W^{\pm}$ gauge Boson mass, $m_W$, predicted by EW-WS theory.

Let me now suppose that another complex singlet Higgs field, $\Psi_{g-EW}(x)$, intervenes to give a mass, $m_g$, to the massless photon field alone, $A^{\mu}$, through SSB of the remaining unbroken $U(1)_{EM}$ gauge symmetry of EW-WS theory. Namely, following the derivation in Sec. 18.2 of Ref. \onlinecite{Mandl}, for example, I can use the results of the original Higgs model with $2+2$ degrees of freedom (dof), which arise from the complex Higgs singlet, $\Psi_{g-EW}(x)$, plus the real originally massless gauge Boson, $A^\mu$. Then another Higgs Boson mass, $m_{Hg}$, results from this second SSB. Of course, one could hardly determine what that additional mass, $m_{Hg}$, ought to be. Notwithstanding that, let me venture to assume that $m_{Hg}$ may be of the order of the already established doublet Higgs Boson mass, $m_H$, hence, $m_{Hg} \sim 125 GeV/c^2$ as well. I supposed that if  $m_{Hg}$ were much lower than $m_H$, it would have been likely detected before. General theoretical constraints further exclude much heavier Higgs masses.\cite{Mandl, peskin1995introduction}

Now the only gauge Boson, $A^{\mu}$, acquires a mass
\begin{equation}\label{mg}
m_g c^2= |e| v_g \sim M_g c^2 \simeq 1.3 \mathrm{x}10^{-41} GeV  , 
\end{equation}  
where in the second step I have introduced my `prescription' of endowing the $g$-photon with a cosmological mass of the order of $M_g$, as given in the fundamental \Eq{massfundamental}.

That yields
\begin{equation}\label{VEVmg}
v_g \sim M_g c^2 / |e| = \sqrt{ h c \Lambda/ 2 \alpha} \simeq 3 \mathrm{x} 10^{-34} \sqrt{GeV/cm}  , 
\end{equation}  
where $h c = 1.2398 \mathrm{x} 10^{-13} GeV \mathrm{x} cm$. This expresses the VEV of the singlet Higgs field, $\Psi_{g-EW}(x)$, as a square-root ratio of fundamental $g$-constants of QM and GR, i.e., $h$, $c$, $\Lambda$, with the QED coupling constant, $|e| = (4 \pi \alpha \hbar c)^{1/2}$. Whatever that may or may not mean, fundamentally, it confirms that a photon mass, $m_g$, can hardly be `tuned' to zero without requiring that $h \rightarrow 0$ is also `tuned' to zero, which is unacceptable.

Maintaining roughly the same value for $\sqrt{-2 \mu^2} \sim m_{Hg} \sim 125 GeV/c^2 $ as in \Eq{EWWS}, I thus obtain that 
\begin{equation}\label{gLambda}
\lambda_g =  \frac{- 2 \mu^2 c^4}{2 \hbar c v_g^2} \sim\frac{- 4 \pi \alpha \mu^2 c^2 }{h^2 \Lambda} \sim 4.4 \mathrm{x}10^{84} .
\end{equation}  
Compared with $\lambda$ in \Eq{EWLambda} for the standard EW-WS theory, I have thus arrived to an `astronomically' larger quartic coupling, $\lambda_g$, for my additional singlet Higgs field. That follows directly from $\Lambda$ being so small in \Eq{fundamental}. However, $\Lambda$ has physical dimensions and it can hardly be `tuned' to zero. Thus $\lambda_g$ in \Eq{gLambda} cannot be let to diverge either.
%
%
%
%
So, perturbative theories or calculations can hardly apply with such a large $\lambda_g$-value. However, non-perturbative theories may still be handled either theoretically or computationally, as in fact they already are in SM constructions.\cite{Nir}

Considering famous `Mexican-hat' pictures such as Fig. 18.1 on p. 406 of Ref. \onlinecite{Mandl}, the True Vacuum drops from the False Vacuum of EW-WS theory about $(v/v_g)^2 \sim 3.6 \mathrm{x}10^{85}$ times more than the True $g$-Vacuum dips from the False $g$-Vacuum of $g$-theory. Likewise, the distance of the True Vacuum plunge from the False Vacuum hump of EW-WS theory exceeds that of the tiny dip of the True $g$-Vacuum from the False $g$-Vacuum tiny bump of $g$-theory by about $v/v_g \sim 6 \mathrm{x}10^{42}$ times.

Thus, I obtain that
\begin{equation}\label{VEVratio}
v_g/v \sim \frac{M_g}{2 sin\theta_W m_W} \simeq 1.7 \mathrm{x}10^{-43} ,  
\end{equation}  
which corresponds to the alternative definition and $\epsilon /g$ value of \Eq{epsilonmass} for the Stuekelberg-Higgs model, as I anticipated. 

This results mainly from the fact that both KMcK theory and my Higgs-Higgs basic approximation essentially decouple the SU(2) Higgs sector from the extant U(1) gauge invariance, which is then subjected to SSB independently. These two theories are thus basically equivalent, hence, virtually `exact,' based on that assumption of independence. Namely, given that with my cosmological `prescription' the ratio $M_g/m_W \sim 1.6 \mathrm{x}10^{-43}$ is so tiny in both \Eq{epsilonmass} and \Eq{VEVratio}, both approximate theories become essentially the same and correspondingly `exact.'

\section{Conclusion}\label{Conclusion}

I considered electro-weak masses and interactions generated for photons by vacuum expectation values of Stueckelberg and Higgs fields. I provided a `prescription' to relate their parametric values to a cosmological range derived from a fundamental Heisenberg uncertainty principle and Einstein-de Sitter cosmological constant and horizon. This yields at least qualitative connections between microscopic ranges acquired by $W^{\pm}$ or $Z^0$ gauge Bosons and the cosmological scale and minimal mass acquired by $g$-photons. I fully applied that procedure to a well established Stueckelberg-Higgs mechanism, i.e., the Kuzmin-McKeon theory developed in Ref. \onlinecite{Kuzmin:2001pg}. That Stueckelberg-Higgs theory, including my `prescription,' may be regarded as a self-contained and complete model of a cosmologically massive photon field, $A^\mu$, with a mass $m_A \sim M_g$, within an order of magnitude or so on the scale of de Sitter horizon. Then I developed a similar, if not equivalent, procedure for a pair of Higgs fields that sequentially or independently break both SU(2) and U(1) gauge invariances of electro-weak interactions. Notwithstanding their smallness, signatures of photon masses and their additional parity-breaking interactions with leptons and neutrinos may be detected in suitable colliders that naturally hold vast amounts of $g$-BEC photonic matter. Furthermore, $m_g$-effects may be observed astronomically through BEC of $g$-photons that may contribute to dark matter and dark energy. My `procedure' and $g$-theory may also justify or further improve previous cosmological estimates of lepton and neutrino masses. With regard to those masses, see also the following Appendix A.

\section{Appendix A: Mass Scale Laws}\label{Appendix A}

After I completed my work and paper, I read a remarkable compendium of Chavanis that further unifies most, if not all known or proposed mass scale laws connecting cosmophysics to microphysics.\cite{ChavanisMassScale} That begins with a deeper analysis of Eddington's relations, yielding \Eq{massfinestructureChavanis} for the electron mass, $m_e$, and fine structure constants, $\alpha$, as previously derived in a logotropic model.\cite{ChavanisLogotropic} Chavanis then goes far beyond that, by deeply analyzing and relating most relevant references, in fact 126 of them, quoted in Ref. \onlinecite{ChavanisMassScale}. 

Based on a minimal fundamental a-dimensional ratio that I also considered in Eqs. [49-52] of Ref. \onlinecite{resca2023cosmological}, i.e.,
\begin{equation}\label{fundamentalratio}
\sqrt \frac{h G}{c^3}  \sqrt \Lambda = \frac{l_P}{L_g} =  \frac{t_P}{T_g} = \frac{M_g}{m_P} \simeq 4.2 \mathrm{x} 10^{-61} \sim \chi^{-1/2} ,
\end{equation}  
I can relate my results to the Mass Scale Law that Chavanis mostly considers, i.e., $\chi^{a/6}$. 

Chavanis is then able to relate parameter values of $a = 3, 2, 1, 0, -1, -2, -3$ to the universe as a whole, Fermion stars, mini Boson stars, Planck black holes, the electron, the neutrino, and the cosmon, respectively. Exceptionally, the neutrino mass that we previously found in \Eq{geometric} corresponds to $a = -3/2$ in Eq. [70] of Ref. \onlinecite{ChavanisMassScale}. However, another type of neutrino may still correspond to the expected $a = -2$ in the Mass Scale Law of Ref. \onlinecite{ChavanisMassScale}. Major progress in the work that I propose may also contribute to clarifying cosmological origins of neutrino masses, as well as that of $g$-photons.  



\end{document}